\documentstyle[12pt,aasms4,flushrt]{article}

\newcommand\psr{PSR~J1617$-$5055}

\newcommand\snr{RCW~103}
\def\lapp{\ifmmode\stackrel{<}{_{\sim}}\else$\stackrel{<}{_{\sim}}$\fi}
\def\gapp{\ifmmode\stackrel{>}{_{\sim}}\else$\stackrel{>}{_{\sim}}$\fi}

\begin{document}

\title{The 69 ms Radio Pulsar Near the Supernova Remnant RCW~103}

\author{V. M. Kaspi and F. Crawford}
\affil{Massachusetts Institute of Technology, Physics Department,
Center for Space Research, 70 Vassar Street, Cambridge, MA 02139}

\author{R. N. Manchester}
\affil{Australia Telescope National Facility, CSIRO, P.O. Box 76,
Epping, NSW 2121, Australia}

\author{A. G. Lyne and F. Camilo}
\affil{University of Manchester, Jodrell Bank, Macclesfield, Cheshire, SK11 9DL, UK}

\author{N. D'Amico}
\affil{Osservatorio Astronomico di Bologna, via Zamboni 33, 40126 Bologna, Italy}

\author{B. M. Gaensler}
\affil{Australia Telescope National Facility, CSIRO, P.O. Box 76,
Epping, NSW 2121, Australia;  Astrophysics Department, School of
Physics, University of Sydney, NSW 2006, Australia}

\begin{abstract}

We report the detection of the radio pulsar counterpart to the 69~ms
X-ray pulsar discovered near
the supernova remnant \snr\ (G332.4$-$0.4).  Our detection confirms
that the pulsations arise from a rotation-powered neutron star, which
we name \psr.  The observed barycentric period derivative
confirms that the pulsar has a
characteristic age of only $8$~kyr, the sixth smallest of all known
pulsars.  The unusual apparent youth of the pulsar and its proximity
to a young remnant
requires that an association be considered.  Although the respective
ages and distances are consistent within substantial uncertainties,
the large inferred pulsar transverse velocity is difficult to explain
given the observed pulsar velocity distribution, the absence of
evidence for a pulsar wind nebula, and the symmetry of the remnant.
Rather, we argue that the objects are likely superposed on the
sky; this is reasonable given the complex area.  
Without an association, the question of where is the
supernova remnant left behind following the birth of PSR~J1617$-$5055
remains open. We
also discuss a possible association between \psr\ and the $\gamma$-ray
source 2CG~333+01.  Though an association is energetically plausible,
it is unlikely given that {\it EGRET} did not detect 2CG~333+01.

\end{abstract}


\section{Introduction}

\snr\ (G332.4$-$0.4) is a young shell supernova remnant with a
complicated observational history.  An X-ray point source near the
center of the remnant was discovered by Tuohy \& Garmire
(1980),\nocite{tg80} who suggested it is a ``radio-quiet''
thermally cooling neutron
star, the stellar remnant of the supernova explosion.  This was
supported by Tuohy et al. (1983) \nocite{tgmd83} and later by
Manchester,
D'Amico \& Tuohy (1985) \nocite{mdt85} and Kaspi et
al. (1996) \nocite{kmj+96} who failed to find evidence for pulsed
emission in radio searches.  Aoki et al. (1992) \nocite{adm92}
detected 69~ms pulsations from the direction of the remnant using {\it
Ginga}, providing additional evidence for the neutron star
identification.  However, {\it Ginga}'s poor spatial resolution ($\sim
1^{\circ} \times 2^{\circ}$) precluded any firm association.  
Recently, Gotthelf,
Petre \& Hwang (1997)\nocite{gph97} detected the central source 
using the {\it ASCA} X-ray
observatory and argued that its spectrum
is harder than that of a cooling neutron star. 
They found no evidence
for pulsations, setting upper limits that were inconsistent with the
Aoki et al. claim.  This history recently took a dramatic turn with
the detection of the 69~ms periodicity in the {\it ASCA} data from
a point source, AXS J161730$-$505505, located $7^{\prime}$ north of
the center of the remnant, outside the 5$^{\prime}$ remnant radius
(\cite{gph97,tkt+98}).  Here we report the discovery of pulsed
radio emission from this pulsar, which we name PSR~J1617$-$5055.

\section{Observations} 

We observed the X-ray pulsar position reported by Gotthelf et al. (1997)
(J2000 RA: 16$^{\rm h}$~17$^{\rm m}$~30$^{\rm s}$, DEC:
$-$50$^{\circ}$~55$'$~05$''$) at the Parkes Observatory 64~m radio
telescope in New South Wales, Australia on 1998 January 15 and 16, for
2.6 and 4.0~h, respectively.  For these observations, we used the
center beam of the Parkes multibeam receiver system at a central radio
frequency of 1374~MHz.  The cryogenically cooled system receives
orthogonal linear polarizations, each of which is down-converted to an
intermediate frequency and filtered in a 2~$\times$~96~$\times$~3~MHz
analog filter-bank spectrometer.  Data were one-bit sampled at
250~$\mu$s and recorded onto magnetic tape using a DLT~7000 tape
recorder attached to a DEC Alpha workstation.  The data acquisition
software and hardware are those being used in a major survey of the
Galactic Plane for pulsars (see \cite{clb+98}).

\begin{figure}
\plotfiddle{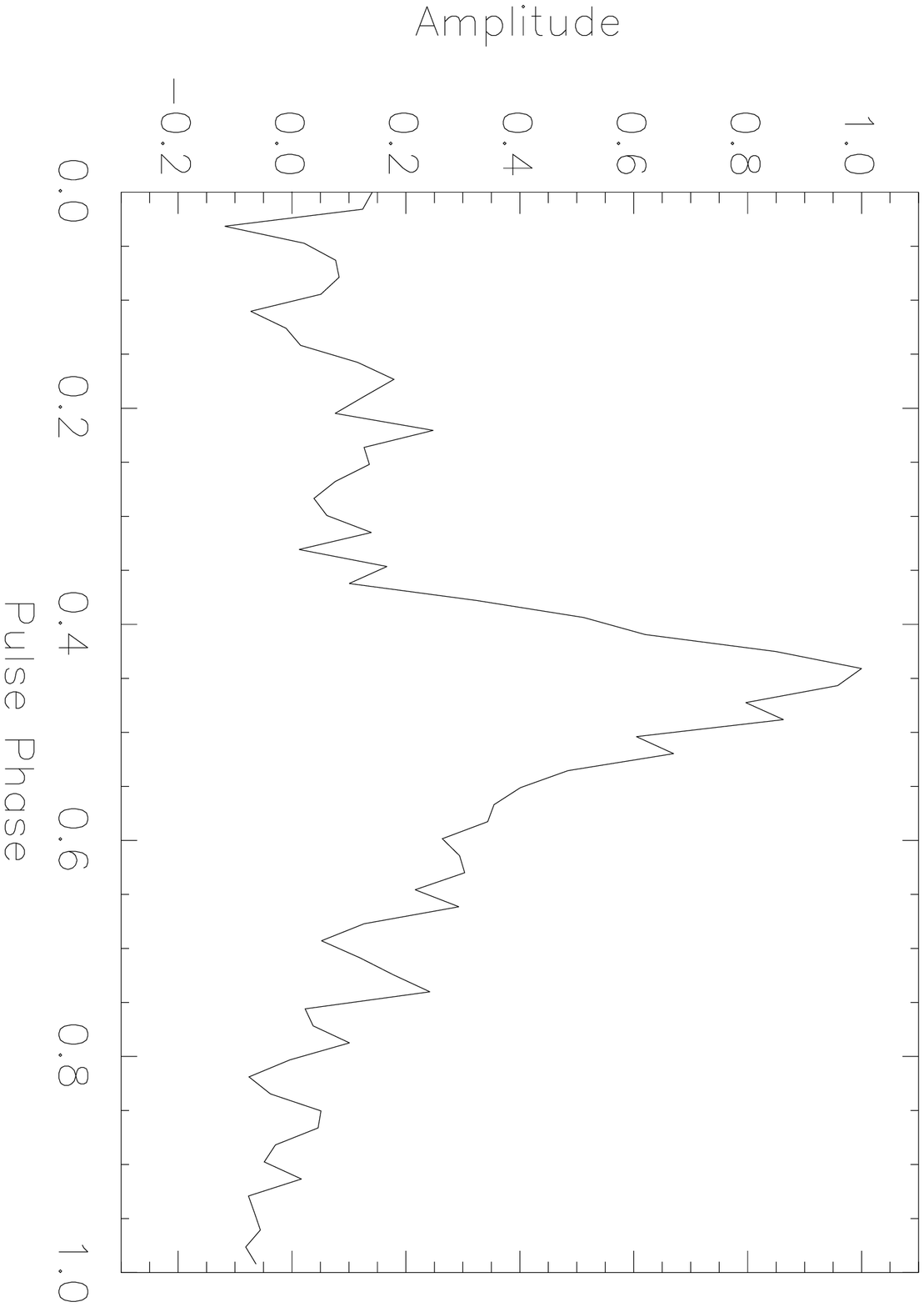}{8in}{90}{75}{75}{250}{0}
\caption{Pulse profile of PSR~J1617$-$5055 at 1.4~GHz.}
\label{fig:prof} 
\end{figure}

In offline processing, the data were dedispersed at dispersion measures
(DMs) between 0 and 600~pc~cm$^{-3}$.  Each dedispersed time series was
folded over a range of topocentric periods $\pm 2$~$\mu$s from that
predicted by the ephemeris estimated by Torii et al. (1998).  A highly
significant detection (signal-to-noise ratio 21) was found at DM = $467
\pm 5$~pc~cm$^{-3}$, at barycentric period
(0.069356847$\pm$0.000000003)~s (epoch MJD 50829.7).  The pulsation was
unambiguously confirmed the next day (signal-to-noise ratio 33).  The
profile obtained by folding the data from the latter observation is
shown in Figure~\ref{fig:prof}.

The 1.4~GHz average radio pulse is unexceptional, characterized by a single
peak of 50\% intensity width $5.8 \pm 0.6$~ms (10\% intensity width $11
\pm 1$~ms), convolved with a one-sided exponential of decay time
constant $8.7 \pm 0.5$~ms, probably due to scattering.  Although the
observations were not carefully calibrated, we estimate the flux
density of the source to be $\sim 0.5$~mJy, with an uncertainty of
$\sim 30$\%.
Using archival Australia Telescope Compact Array (ATCA) data taken of
the area, but filtering out large-scale structure,
seven unresolved sources can be seen within $\sim$4$'$ of the pulsar
which have 1.4~GHz flux densities between 0.5 and 1.4 mJy.  None of
these sources is noticeably circularly or linearly polarized, 
although bandwidth depolarization could easily account for
the latter possibility.  The source closest to the {\it ASCA} position
has flux density $\sim$0.5~mJy, consistent with our
estimate from Parkes and a flat spectrum similar to other young
pulsars: it is at J2000 RA: 16$^{\rm h}$~17$^{\rm m}$~29.3$^{\rm s}$,
DEC: $-$50$^{\circ}$~55$'$~13.2$''$ (uncertainty $\sim$0.2$''$).

We searched for occurrences of giant radio pulses from PSR
J1617$-$5055 by combining individual dedispersed samples to form a 20 ms
resolution time series.
Significantly narrower pulses were unlikely to be seen since dispersion
smearing and multipath scattering should combine to broaden pulses by
$\sim$10 ms.  We found no occurrences of single pulses having amplitude
exceeding 9 times the RMS noise, corresponding to 54
times the mean pulse energy.  For an emission rate and amplitude
distribution of giant pulses as observed for the Crab pulsar
(\cite{ag72,lcu+95}), we would expect $\gapp$5 occurrences above this
energy threshold in our data.  Our analysis thus demonstrates that
PSR~1617$-$5055 is not emitting Crab-like giant radio pulses.

It is straightforward to understand why previous searches for radio
pulsations from \snr\ at Parkes 
did not detect \psr.  They were conducted at radio frequencies near 430,
660 and 1520~MHz: at the two lower frequencies, interstellar
scattering is expected to have broadened the pulse beyond
detectability.  At 1520~MHz, the Parkes telescope beam has FWHM
$\sim$13$'$, so the sensitivity to \psr\ is reduced by $\gapp$50\%
when pointing at the center of \snr.

\section{Discussion}

Using the barycentric period of our radio detection and that
found by Torii et al. (1998), we find that $\dot{P}=1.351(2) \times
10^{-13}$, consistent with the value of $1.4 \times 10^{-13}$ that
Torii et al. derive using the {\it Ginga} detection.\footnote{The
uncertainty in $\dot{P}$ derived from the {\it Ginga} period is at
least $0.1 \times 10^{-13}$ and probably larger.  It is dominated by
the uncertainty in the {\it Ginga} period, which was not quoted by
Aoki et al. (1992).}  From the measured $\dot{P}$ we infer a surface
magnetic field strength $B \equiv 3.2 \times 10^{19} (P \dot{P})^{1/2}
= 3.1 \times 10^{12}$~G, and a spin-down luminosity $\dot{E} \equiv 4
\pi^2 I \dot{P} / P^3 = 1.6 \times 10^{37}$~erg~s$^{-1}$, where $I$ is
the neutron star moment of inertia, assumed to be $10^{45}$~g~cm$^2$.
The implied characteristic age for \psr\ is $\tau_{c} \equiv
P/2\dot{P} = 8.1$~kyr. This is the sixth smallest known pulsar
characteristic age after the Crab pulsar, PSRs B0540$-$69, B1509$-$58,
B1610$-$50 and the recently discovered J0537$-$6910 (\cite{mgz+98}).

The apparent proximity of so young a pulsar to a young supernova
remnant demands that an association be considered.  If they are
associated, the nature of the well-studied central X-ray source would
be unclear, as no tenable interpretations other than its being an
isolated cooling neutron star have been put forward
(\cite{pop98,hh98}).  The probability of chance superposition of the
pulsar near the remnant is difficult to quantify but is certainly not
negligible, particularly in this complex region of the sky: this
line-of-sight traverses the Sagittarius-Carina and Scutum-Crux spiral
arms and may extend to the Norma arm.  Indeed, the very young radio
pulsar PSR~B1610$-$50, only $\sim$30$'$ away on the sky from \psr, is
also probably superposed only by chance near the supernova remnant Kes~32
(\cite{jml+95,gj95c}, but see also \cite{car93}).  We now consider the
evidence for an association between \psr\ and \snr.

\subsection{Do independent distance estimates for the pulsar and
remnant agree?}
\label{sec:dist}

First we consider the remnant distance.  Westerlund (1969)
\nocite{wes69} estimated the distance to \snr\ to be
$d=3.9$~kpc, by associating it with OB stars nearby on the sky.
Caswell et al. (1975) \nocite{cmr+75} used HI absorption to establish
a systemic velocity of $-$44~km~s$^{-1}$ for the remnant, which, using
the rotation curve of Fich, Blitz \& Stark (1989)
\nocite{fbs89} and standard IAU parameters
(\cite{klb86}), corresponds to a distance of $3.1 \pm 0.4$~kpc.  This
has often been taken to be the actual distance to the source
(e.g. \cite{dgym96,gph97}); however, the signal-to-noise in the
absorption spectrum is low and 3.1~kpc should more reasonably be
adopted as a lower limit on the distance.
Leibowitz \& Danziger (1983) and Ruiz (1983) \nocite{ld83,rui83}
independently estimated the distance to be $\sim 6.5$~kpc from the
visual extinction of optical filaments.  All observations are thus
reconciled if $d\sim 6.5$~kpc, and the remnant is not associated with
Westerlund's OB stars.

Next we consider the pulsar distance.  Using the standard DM-distance
model (\cite{tc93}), the observed pulsar DM implies a distance of
6.1--6.9~kpc.  This close agreement with the \snr\ distance estimated
above could be merely fortuitous.  A comparison of pulsar distances
from HI absorption and the DM-distance model reveals that the latter
may systematically underestimate the electron density for pulsars near
\psr.  There are two pulsars within 20$^{\circ}$ of \psr\ that have
similar DMs and have distance estimates from HI absorption.
PSR~B1641$-$45 has DM=475~pc~cm$^{-3}$ and HI absorption lower and
upper distance limits of 4.2$\pm$0.3 and 5.0$\pm$0.3~kpc, respectively
(\cite{fw90}), while the Taylor \& Cordes (1993) model predicts
5.7--6.4~kpc.  Similarly, PSR~B1718$-$35 (DM=496~pc~cm$^{-3}$) has an
HI distance range of 4.4$\pm$0.5 to 5.2$\pm$0.6~kpc (\cite{wsfj95}),
while the DM-distance model reports 5.8--7.5~kpc.  If we adopt the
mean electron density for these two pulsars for \psr, its distance may
be as low as 4.5~kpc.  There is no evidence for any line-of-sight HII
region that could significantly contribute to the DM of \psr.

The X-ray luminosity $L_x$ of plerionic nebulae
powered by pulsars is correlated with spin-down luminosity $\dot{E}$
(\cite{sw88,bt97}).  Torii et al. (1998) \nocite{tkt+98} argue
that if the unpulsed component of the X-ray emission from the
direction of \psr\ is from an unresolved synchrotron nebula, then its
observed $L_x$ is consistent with the range predicted by empirical
$L_x / \dot{E}$ relationships only if the pulsar's distance is much
larger than 3.1~kpc.  For $d\sim 6.5$~kpc, the observed $L_x$ is still
somewhat lower than that predicted by the empirical relationship, but
is at least within the scatter delimited by other sources.

Thus, distance estimates to \psr\ and \snr\ agree,
though both have sufficiently large uncertainty that the agreement
is not strong evidence for an association.  

\subsection{Do independent age estimates for the pulsar and remnant
agree?}

\snr\ has long been thought to be a very young remnant because of its
symmetric shell morphology, which suggests that it has had insufficient
time to be distorted by irregularities in the ambient medium.  The
X-ray structure supports this view, as it is characteristic of
the young ``double-shock'' evolutionary stage (\cite{che82}).  However,
the observed correlation of the optical and infrared line emission
with the radio filaments suggests that the remnant has reached the
Sedov point-blast stage.  This is supported by the remnant's
polarization structure, which indicates an absence of the radial
magnetic field pattern seen in all young double-shock shell remnants
(\cite{dgym96}).  On this basis, Dickel et al. suggest that the
remnant has just entered the Sedov stage.  Assuming a distance of
3.1~kpc, given the observed angular diameter, Dickel et al. argue that
the remnant probably has an age of $\sim$1~kyr.  Using the same arguments
but taking the distance to be 6.5~kpc (\S\ref{sec:dist}), the age is
$\sim$2~kyr.  A larger age has been estimated from X-ray observations.
Gotthelf et al. (1997) \nocite{gph97} studied the remnant's X-ray
spectrum, and, using a non-equilibrium ionization plasma model for
Sedov hydrodynamics (\cite{hsc83}), deduced an age of 4~kyr, for a
distance of 3.1~kpc.  The larger distance would correspondingly
increase this age estimate.  However, the most direct method of age
estimation comes from Carter, Dickel \& Bomans (1997) \nocite{cdb97} who have
detected the mean expansion rate of the remnant to be $1''.8 \pm
0''.2$ per 25~yr using optical images taken 25~yr apart.  They
conclude that the remnant can be no older than $\sim$3~kyr, and is
most likely 2~kyr old.  These estimates do not depend on
the distance to the source.

\begin{figure}
\plotone{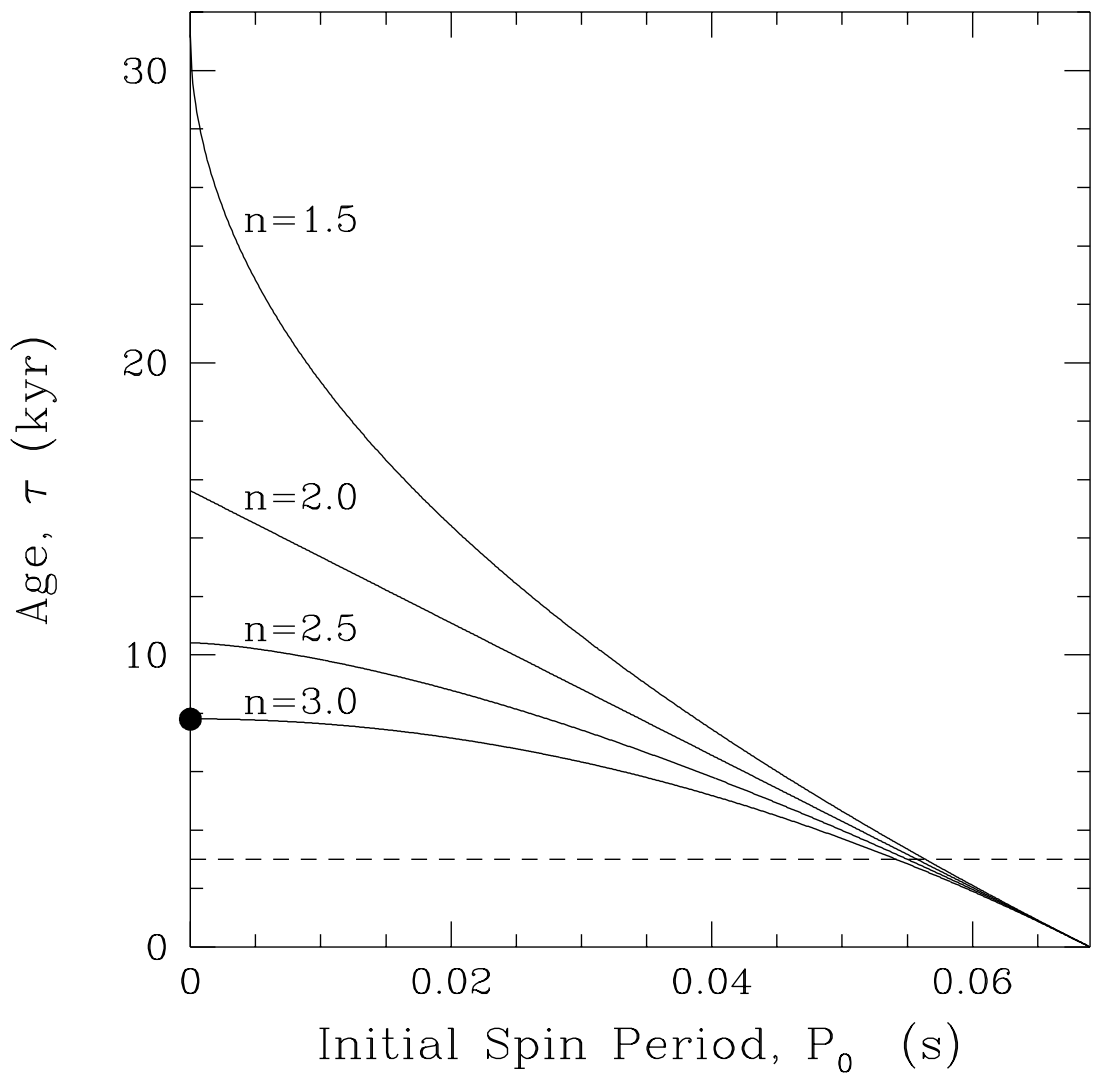}
\caption{The true age $\tau$ of \psr\ as a function of the assumed initial spin
period $P_0$
for four values of the braking index $n$.  The dashed line represents the
estimated upper bound on the age of \snr.  The filled circle represents
the characteristic age  $\tau_c \equiv P/2\dot{P}$ (see text).}
\label{fig:spin} 
\end{figure}

The unusually low characteristic age $\tau_c = 8$~kyr for the pulsar
tends to support an association with \snr.  However, the
characteristic age reflects the true pulsar age only if the braking
index $n=3$, as expected for a simple magnetic-dipole braking, and if
the initial spin period $P_0$ is much less than the current period.
Figure~\ref{fig:spin} summarizes the dependence of the age on these
two parameters.  $P_0$, an unknown, is plotted along the x-axis, and
the true pulsar age $\tau$ on the y-axis.  The curves show the
dependence of the true age on $P_0$ for several $n$ that span the
range of observed values.  Note that for $P_0 < 20$~ms and $n<3$, as
is the case for those pulsars for which these quantities are known
with certainty (a particularly low value of $n$ has recently been
reported for the Vela pulsar -- see
\cite{lpsc96}), the true age of \psr\ must
be considerably greater than that inferred for \snr.  The horizontal
dashed line represents the estimated upper bound on the remnant age,
$\sim 3$~kyr.  The intersection of the dashed line with each curve
gives $P_0$ for each $n$.  Thus, if the pulsar and remnant are
associated, then $P_0$ is $\sim$50~ms, independent of $n$.  This value
for $P_0$ would not be surprising given the range implied by the range
of current spin periods for the very youngest pulsars (e.g. PSR
B1509$-$58 has $P=150$~ms).

\subsection{Is the implied pulsar transverse velocity reasonable?}
\label{sec:vel}

If \psr\ and \snr\ are associated, the pulsar birth place must be
near the remnant center.  The symmetry of the shell makes
the birth place easy to identify, in contrast to other proposed
associations, such as that of PSR~B1509$-$58.  For an association, the
implied proper motion of \psr\ would be
$\sim$130~(3~kyr/$\tau$)~mas~yr$^{-1}$, and the corresponding
transverse velocity is
$\sim$4200~($d$/6.5~kpc)~(3~kyr/$\tau$)~km~s$^{-1}$.  This velocity
would be unusually large given the observed pulsar velocity
distribution (\cite{ll94,hp97}).  Simulations of remnant
evolution assuming the Lyne \& Lorimer (1994) 
velocity distribution
suggest that $<$2\% of young pulsars
should have left their parent shells (\cite{gj95b}).
If an association exists, this pulsar would easily rank among the
fastest-moving stellar objects in the Galaxy and would have important
implications for models of supernova explosions (c.f. \cite{bh96b}) and
binary evolution (e.g. \cite{bp95,fk97}).  The large required space
motion makes it easy to verify observationally; the proper
motion may be detectable from radio timing observations in a few years
(although frequent glitches, as expected from this young
pulsar, could preclude such a measurement); similarly, gated radio
imaging and high spatial resolution X-ray observations over
several years should detect the motion.

\subsection{Is there evidence for a pulsar wind nebula?}

For an association, the required high pulsar velocity suggests that a
pulsar wind nebula might be observable, a result of the confinement of
the pulsar's relativistic wind by ram pressure (see \cite{cor96} for a review).
The spectacular nebula observed for the young, energetic radio pulsar
PSR~B1757$-$24, apparently located just outside the supernova
remnant G5.4$-$1.2, is strong evidence for a high pulsar velocity and
hence the association with the remnant (\cite{fk91,mkj+91}).  A
similar nebula might be expected near \psr, if indeed it has a high
space velocity.  Dickel et al. (1996) \nocite{dgym96} published
high-resolution radio maps of \snr\ that just fail to include the position of
\psr.  A re-analysis of their ATCA
1.4~GHz data shows no evidence in the pulsar direction for extended
emission on spatial scales between 6$''$ and 30$'$, down to a
3$\sigma$ limit of 1.2~mJy~beam$^{-1}$.  A nebula like that observed
for PSR~B1757$-$24 would have been detectable.  This provides evidence
against a high velocity, particularly because \psr\ has spin-down luminosity
$\dot{E}$ an order of magnitude larger than that of PSR~B1757$-$24.
Brighter nebulae are expected for larger ambient densities, although the
latter is estimated to be very small for PSR~B1757$-$24.  Furthermore,
there is no evidence from the morphology of \snr\ that an energetic
pulsar passed through its shell. The spectacular shell ``rejuvenation''
interactions observed
for PSR~B1757$-$24/G5.4$-$1.2 and, for example, PSR~B1951+32/CTB~80
(\cite{fss88,sfs89,hk88}) are absent in \snr.  This is evidence against an
association, particularly given how recently the pulsar would have to
have crossed the shell.

\section{\psr:  A $\gamma$-ray pulsar?}

High-energy $\gamma$-ray emission from rotation-powered
pulsars is observed to be well correlated with spin-down luminosity
corrected for distance (see \cite{tho96b} for a review).
Assuming $d\sim 6.5$~kpc for \psr, its $\dot{E}/d^2$ is over 30 times
larger than that for PSR~B1055$-$52, a known $\gamma$-ray pulsar.
\psr\ is therefore an excellent candidate for observable $\gamma$-ray
emission.  Indeed the {\it ASCA} position of \psr\ lies $0.6^{\circ}$
outside the 90\% confidence $1^{\circ}$ error radius of the
high-energy $\gamma$-ray source 2CG~333+01 (\cite{sbb+81}).  If
2CG~333+01 is the counterpart of \psr, then the conversion efficiency
of $\dot{E}$ to $E>100$~MeV $\gamma$-rays is $\sim 0.1(d/6.5 \; {\rm
kpc})^2$, assuming 1~sr beaming and a differential $\gamma$-ray photon
index of 2.  This efficiency is comparable to those of other known
$\gamma$-ray pulsars.  However, a counterpart to 2CG~333+01 appears neither
in the Second {\it EGRET} catalog (\cite{tbd+95}), nor in the
Lamb \& Macomb (1997) \nocite{lm97} catalog of GeV {\it EGRET}
sources.  
As $\gamma$-ray fluxes of pulsars are
observed to be steady (\cite{mmct96}), we conclude that unless \psr\ is the first
counter-example, or the absence of an {\it EGRET} counterpart can be
explained otherwise, the association of \psr\ and 2CG~333+01 is doubtful.

\section{Conclusions}

Our discovery of radio pulsations from \psr\ confirms that the 69~ms
pulsations observed by Torii et al. (1998) come from a young,
energetic rotation-powered pulsar in the direction of the young
supernova remnant RCW~103.  We have considered the possible
association between \psr\ and \snr\ and argue that it is unlikely,
given the required transverse velocity, which is much higher than that
implied by the observed pulsar velocity distribution, and the absence
of evidence for a pulsar-powered nebula or any effect on the remnant shell.
This same large velocity makes our conclusion easy to test
observationally, as the corresponding large proper motion should be easily
detectable.  In the absence of an association, the interesting
question of where is the supernova remnant left behind following
the birth of the very young PSR~J1617$-$5055 is (similarly for 
the young PSR~B1610$-$50, also not clearly 
associated with a remnant), remains open.  One possibility is
that their braking indexes are  smaller than the canonical 3.0,
as in the older Vela pulsar, rendering their characteristic ages
inappropriately small and  giving more time for the remnant to have
faded from view.  Long-term timing observations can test this hypothesis.
Finally, we note that the high $\dot{E}$ for \psr\
implies that it should be an observable $\gamma$-ray pulsar.  The
pulsar's coincidence with an unidentified COS~B source 2CG~333+01 is
interesting, but the absence of any counterpart detected by {\it
EGRET} is problematic.

\acknowledgments

We are deeply indebted to Ken'ichi Torii for communicating to us
information about the {\it ASCA} discovery before publication.
We thank E. Gotthelf and N. Kawai for valuable discussions
and J. Dickel and A. Green for providing their ATCA data of the
region.  The Parkes radio telescope forms part of the Australia
Telescope, which is funded by the Commonwealth of Australia for
operation as a National Facility operated by CSIRO.
FC acknowledges support from the European Commission
through a Marie Curie fellowship under contract no. ERB FMB1 CT961700.

\newpage


\clearpage

\end{document}